\documentclass[twocolumn,prx,english,superscriptaddress]{revtex4-2}
\usepackage{amsfonts}
 \usepackage{amsmath}
\usepackage{amssymb}
\usepackage[colorlinks, citecolor=blue,linkcolor=red]{hyperref}
\usepackage{color}
\usepackage{float}
 \usepackage{textcomp}
 \usepackage{subfigure}
 \usepackage[rightcaption]{sidecap}
 \usepackage{graphicx}
\usepackage{braket}
\usepackage{xcolor}
\usepackage[normalem]{ulem}

\newcommand{\reduline}[1]{\bgroup\color{red}\uline{\textcolor{black}{#1}}\egroup}

\begin{document}
\title{Emergent Incompressibility in Polariton Condensates for Scalable XY Spin Simulation}

\author{Chengyong Yu}
\affiliation{School of Science and Engineering, The Chinese University of Hong Kong (Shenzhen), Longgang, Shenzhen, Guangdong, 518172, P.R. China}

\author{Youya Xu}
\affiliation{School of Science and Engineering, The Chinese University of Hong Kong (Shenzhen), Longgang, Shenzhen, Guangdong, 518172, P.R. China}

\author{Xuan Ye}
\affiliation{School of Science and Engineering, The Chinese University of Hong Kong (Shenzhen), Longgang, Shenzhen, Guangdong, 518172, P.R. China}

\author{Yan Zhou}
 \affiliation{School of Science and Engineering, The Chinese University of Hong Kong (Shenzhen), Longgang, Shenzhen, Guangdong, 518172, P.R. China}

\author{Sanjib Ghosh}
\email{sanjibghosh@cuhk.edu.cn}
\affiliation{School of Science and Engineering, The Chinese University of Hong Kong (Shenzhen), Longgang, Shenzhen, Guangdong, 518172, P.R. China}
\affiliation{Beijing Academy of Quantum Information Sciences, Beijing, 100193, P. R. China}
\setlength{\parskip}{0pt}
\setlength{\belowcaptionskip}{-10pt}

\begin{abstract}
Driven--dissipative polariton networks offer a powerful analog platform for simulating continuous-spin Hamiltonians. However, their practical scalability is fundamentally limited by amplitude--phase coupling and metastable trapping in highly frustrated energy landscapes. Here, we present an \textit{in situ} dual-enhancement mechanism to address these bottlenecks without requiring external  site-specific fine-tuning. First, a renormalization-group analysis based on density rescaling shows that higher-order nonlinear loss drives the condensates toward an incompressible regime, intrinsically suppressing amplitude fluctuations and enforcing phase-only dynamics. For macroscopic XY networks (up to $10{,}000$ spins), this phase--amplitude decoupling yields a more than $50$-fold reduction in energy, while simultaneously shrinking the standard deviation by a factor of six. More generally, it restores the extensive scaling of the optimized XY energy with system size. Second, to navigate the rugged, nonconvex energy landscape, we introduce weak Hermitian pulses. When implemented through a multistage protocol, these pulses enable repeated escapes from local minima and drive the system to deeper energy minima. Benchmarking shows an increasing advantage with system size, with the polariton simulator outperforming the classical Monte Carlo--SQP baseline in every sampled macroscopic realization. Our work provides a physical framework for scalable polariton analog computing.
\end{abstract}

\date{\today}
\maketitle

\textbf{Introduction:--}
Many optimization tasks of practical and scientific relevance, including combinatorial optimization, neural-network training, graph partitioning, and materials design, involve searching an exponentially large configuration space~\cite{lucas2014ising,rojas2013neural,lokhov2018optimal,inagaki2016coherent,wang2026phase}. Identifying the global optimum for such problems is generally computationally intractable, with many belonging to the class of NP-hard problems~\cite{pardalos1991quadratic}. A broad family of these tasks can be mapped onto the minimization of classical spin Hamiltonians, most notably the Ising and XY models~\cite{berloff2017realizing,PhysRevB.111.014307,luo2023wavelength,sun2025programmable}. Whereas the Ising model contains discrete spin variables, the XY model involves continuous angular degrees of freedom; the general weighted XY objective can be written as a unit-modulus quadratic program, a problem class that is NP-hard in general~\cite{8534389}. In frustrated spin-glass graphs, the Hamiltonian exhibits a rugged energy landscape containing many metastable local minima, making global optimization increasingly challenging for conventional algorithms as the system size grows~\cite{mehta2013potential}.

Driven-dissipative Bose--Einstein condensates provide a natural physical approach to this class of optimization problems~\cite{wouters2007excitations,gross2017quantum}. Their macroscopic phases can encode continuous spin variables, while gain, loss, and inter-site coupling implement an analog search for low-energy configurations through dissipative relaxation~\cite{deng2010exciton,byrnes2014exciton}. Exciton-polariton condensates are particularly attractive because of their strong nonlinear interactions, optical accessibility, and the possibility of engineering flexible graph geometries and coupling patterns~\cite{berloff2017realizing, deng2002condensation,kasprzak2006bose,tao2022halide,PhysRevApplied.22.L021001,cao2025polariton,ExtendedXY2025}. Building on the theory of nonequilibrium condensation, graph simulators have been developed as analog solvers of XY-type Hamiltonians, with phase locking and tunable dissipative coupling demonstrated in experiment~\cite{berloff2017realizing,PhysRevB.101.155402,peng2024room}.

While these studies establish the feasibility of polariton condensates as analog phase simulators, fundamental challenges continue to limit their practical scalability. First, residual phase--amplitude coupling breaks the strict equivalence to a fixed-amplitude spin model. Consequently, the governing cost function of the network depends on both the phase and amplitude degrees of freedom~\cite{PhysRevResearch.2.043335,Gershenzon2020,Verstraelen2024}. 
To mitigate this phase--amplitude crosstalk, prior studies have employed adaptive gain control or auxiliary error variables to externally enforce amplitude homogeneity~\cite{kalinin2018global,PhysRevLett.122.040607}. However, since amplitude variations are dynamically coupled to the relative phase differences, determining the exact gain profile necessary to correct them scales similarly in complexity to solving the underlying XY Hamiltonian. This inherent circularity introduces real-time feedback overhead and can completely negate the potential analog speedup of the polariton simulator in large-scale applications. Second, purely dissipative evolution is locally gradient-like and can settle into metastable attractors rather than the global minimum, a limitation that becomes increasingly severe in large, rugged energy landscapes~\cite{PhysRevResearch.2.043335}.

\begin{figure*}[!htbp]
\centering
\includegraphics[width=1.8\columnwidth]{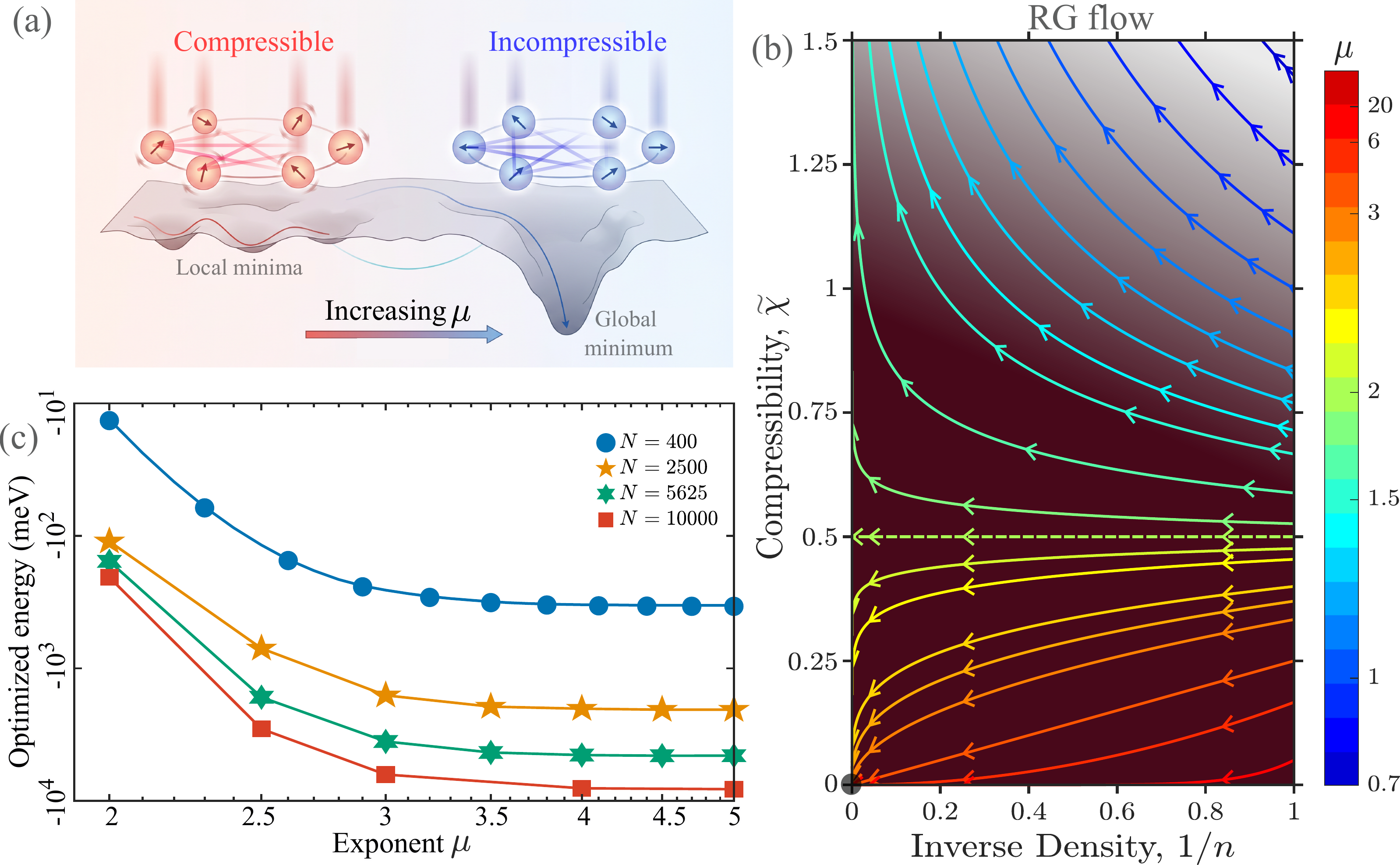}
\caption{\textbf{An incompressible polariton spin simulator}. \textbf{(a)} Optimization dynamics in driven--dissipative polariton spin simulators. In the compressible regime, amplitude fluctuations cause strong phase--amplitude coupling, trapping the system in local minima. A large power exponent $\mu$ suppresses these fluctuations, restoring phase-only dynamics. Coherent Josephson tunneling then facilitates escape to deeper global minima. Sphere size and arrows denote amplitude and phase, respectively. 
\textbf{(b)} Renormalization-group flow of the dimensionless compressibility $\tilde{\chi}$ as a function of inverse density $1/n$. Under density rescaling $n\to bn$ with $b>1$, $\tilde{\chi}$ flows toward the asymptotic limit $\tilde{\chi}\to 0$ for $\mu>2$ (incompressible regime) and increases for $\mu<2$ (compressible regime). The dashed curve marks the critical case $\mu=2$, for which $\tilde{\chi}$ is constant. The colors indicate the value of $\mu$.
\textbf{(c)} Average optimized energy $\langle E\rangle$ as a function of the nonlinear-loss exponent $\mu$ for 2D square lattices with $N$ sites. Increasing $\mu$ from $2$ to $5$ substantially lowers the optimized energy, with the improvement becoming pronounced at larger system sizes (up to a $54$-fold reduction at $N=10,000$). The gradual leveling off of the optimized energy for $\mu\gtrsim 4$ reflects diminishing additional improvement once the system enters the strongly incompressible regime.}
\label{N4IntensityPhase}
\end{figure*}

Here, we investigate the convergence dynamics of driven--dissipative polariton networks to establish a scalable spin simulator. We demonstrate an intrinsic dual-enhancement mechanism that systematically addresses the aforementioned bottlenecks \textit{in situ}, requiring no external intervention. First, through a renormalization-group (RG) flow analysis, we show that higher-order nonlinear loss drives the condensates toward an incompressible scaling asymptote. In this regime, residual amplitude--phase coupling is intrinsically suppressed, effectively restricting the dynamics to a phase-only manifold. Compared to the conventional compressible regime, this decoupling yields substantially deeper optimized energies. Importantly, this energy reduction scales extensively with system size, achieving more than a $50$-fold energy drop for the largest network considered. Second, we introduce weak Hermitian pulsed couplings that operate alongside the primary dissipative dynamics to induce collective coherent tunneling. This coherent contribution facilitates inter-basin mixing among proximal metastable states, providing a physical mechanism to escape rugged local minima.

Finally, we integrate these mechanisms into a multi-stage protocol, applying the Josephson tunneling via a sequence of pulses. This approach enables repeated escapes from metastable traps, consistently driving the system deeper into the optimization landscape. We rigorously evaluate this complete framework on randomly coupled two-dimensional spin-glass lattices, characterized by disordered coupling magnitudes and signs. Ultimately, by unifying macroscopic phase stability with coherent tunneling to solve such highly frustrated systems, our work establishes a robust physical and algorithmic foundation for scalable analog spin simulators.


\textbf{Basic Theory:--} Analog simulation of the classical XY spin Hamiltonian, $H_\mathrm{XY}=-\sum_{k>j}J_{jk} \cos(\theta_k-\theta_j)$ where $\theta_j$ is the continuous spin angle and $J_{jk}$ denotes the coupling matrix, presents a compelling frontier for solving complex combinatorial optimization problems. Mathematically, the ground state of this spin system can be systematically sought via the continuous-time gradient descent method, which relaxes the phase configuration along the local energy gradient according to the dynamical equation: $\dot{\theta}_j = -\partial H_\mathrm{XY}/\partial\theta_j= -\sum_k J_{jk} \sin(\theta_j-\theta_k)$. While executing this strategy computationally becomes intensive for large or highly frustrated networks, it can be physically implemented and accelerated using analog hardware platforms. Driven-dissipative exciton-polariton condensate networks provide an ideal architecture for this purpose, mapping mathematical optimization directly onto physical relaxation dynamics [see Fig.~\ref{N4IntensityPhase}(a)]. In this system, each localized condensate center is characterized by a macroscopic wave function $\psi_j$, with the centers interconnected via incoherent couplings $J_{jk}$~\cite{Kalinin_2018}. The condensate network dynamics are governed by a discrete driven-dissipative effective Gross-Pitaevskii (dGP) theory:
\begin{equation}\label{effective_GPE}
    \hbar\dot{\psi}_j = \sum_{k}J_{jk} \psi_k +  \left( \frac{P - \gamma}{2}- \alpha |\psi_j|^\mu \right) \psi_j -i g |\psi_j|^2 \psi_j,
\end{equation}
where $P$, $\gamma$, $\alpha$, and $g$ represent the non-resonant pump, linear-decay, nonlinear-decay, and polariton nonlinearity, respectively. Here, we introduce an effective exponent $\mu$ for the  nonlinear-decay term that characterizes the density dependence of the nonlinear damping and, consequently, the stiffness of the condensate population. In the standard reservoir model, adiabatic elimination of a weakly depleted excitonic reservoir yields a cubic nonlinear-loss term with $\mu=2$~\cite{wouters2007excitations,RevModPhys.85.299}, for which the steady-state condensate density grows linearly with the net gain, $|\psi|^2\simeq(P-\gamma)/(2\alpha)$. Above threshold, however, reservoir depletion, phase-space filling, and oscillator-strength saturation progressively induce sublinear density growth~\cite{PhysRevB.49.16565,PhysRevB.105.224515,r8m8-hckx}. In the mean-field description, these high-density nonlinear effects can be represented through $\mu>2$, with larger $\mu$ corresponding to a stiffer condensate population. A detailed analysis of $\mu$ is provided in the Supplementary Information.

\textbf{Phase--Amplitude Coupling:--} Applying the Madelung transformation $\psi_j = R_j e^{i\theta_j}$ and considering real-valued coupling constants $J_{jk}$, the dynamics of the condensate phases are governed by:
\begin{equation}\label{coupled_dynamics}
\hbar\dot{\theta}_j = -gR_j^2 - \sum_k \frac{R_k}{R_j} J_{jk}\sin(\theta_{jk}),
\end{equation}
where $\theta_{jk}=\theta_j-\theta_k$. Eq.~\eqref{coupled_dynamics} can implement effective gradient-descent dynamics if the amplitude and phase degrees of freedom are fully decoupled, which requires the amplitudes $R_j$ to remain static. 
In physical networks, however, the amplitudes remain dynamical and are coupled to the phase configuration through a coupling-induced, site-dependent correction to the effective linear gain, defined as $\delta P_j(\boldsymbol{\theta}) \equiv \frac{2}{R_j}\sum_k J_{jk}R_k\cos(\theta_{jk})$, , where $\boldsymbol{\theta}$ denotes the global phase configuration. The corresponding amplitude dynamics are given by:
\begin{equation}\label{coupled_dynamics2}
    \hbar\dot{R}_j = \frac{1}{2}(P + \delta P_j(\boldsymbol{\theta}) - \gamma)R_j - \alpha R_j^{\mu+1}.
\end{equation}
For $(P-\gamma)\gg |\delta P_j|$, one obtains a baseline steady-state amplitude $R_j^2 = n_0 \simeq [(P-\gamma)/(2\alpha)]^{2/\mu}$. However, the inherent phase asymmetry caused by the nonuniform couplings $J_{ij}$ of a spin glass induces local pump fluctuations, $\delta P_j(\boldsymbol{\theta})$. Even small fluctuations around the baseline amplitude can significantly perturb the phase evolution through Eq.~\eqref{coupled_dynamics}. This effect is particularly pronounced in frustrated networks (nonuniform $J_{jk}$), where $\delta P_j(\boldsymbol{\theta})$ remains highly nonuniform even in the ground-state phase configuration.

Eq.~\eqref{coupled_dynamics2} exposes a fundamental challenge in external, feedback-driven amplitude-homogenization schemes~\cite{kalinin2018global,PhysRevLett.122.040607}. As dictated by the dynamics, the amplitude inhomogeneity is generated precisely by the site-dependent gain perturbation $\delta P_j(\boldsymbol{\theta})$, which is determined by the instantaneous phase configuration. Consequently, for an external feedback loop to enforce amplitude homogeneity, it must dynamically evaluate and perfectly neutralize this phase-induced perturbation at every site. However, successfully calculating this exact compensation requires partial knowledge of the optimal spin configuration itself. This inherent circularity defeats the computational advantage of utilizing an analog polariton simulator at large scales. In contrast, our approach retains a strictly uniform spatial pump, intrinsically suppressing the population response to $\delta P_j(\boldsymbol{\theta})$ through the physics of higher-order nonlinear loss.

\textbf{Phase Only Theory:--} To systematically quantify the amplitude fluctuations, we formally construct an effective phase-only Hamiltonian functional, $\mathcal{H}_{\mathrm{eff}}(\boldsymbol{\theta})$, such that the coupled dynamics of Eqs.~\eqref{coupled_dynamics} and \eqref{coupled_dynamics2} can be expressed as:
$\hbar\dot{\theta}_j = -{\partial}\mathcal{H}_{\mathrm{eff}}(\boldsymbol{\theta})/{\partial\theta_j}$.
This represents gradient-descent dynamics over the energy landscape defined by $\mathcal{H}_{\mathrm{eff}}(\boldsymbol{\theta})$. In general, the resulting phase dynamics deviate significantly from our target XY dynamics due to the continuous phase--amplitude feedback loop governed by Eq.~\eqref{coupled_dynamics2}. Consequently, deriving a closed-form analytical expression for a purely phase-dependent Hamiltonian is challenging. 

However, recognizing that Eq.~\eqref{coupled_dynamics2} admits a closed steady-state solution in the absence of the network coupling $J_{jk}$, we can perturbatively solve for the site-dependent density up to second order. Substituting this perturbative steady-state density back into the full Hamiltonian yields the normalized, effective phase-only Hamiltonian (see Supplementary Information for the full derivation):
\begin{equation}
{\mathcal{H}}_{\mathrm{eff}}(\boldsymbol{\theta}) = H_\mathrm{XY} + \chi \mathcal{O}_{\mathrm{pert}}(\boldsymbol{\theta}).
\end{equation}
The corresponding perturbation operator $\mathcal{O}_{\mathrm{pert}}$, which represents the deviation of the effective Hamiltonian from the target XY model, is given by:
\begin{equation}
\mathcal{O}_{\mathrm{pert}}(\boldsymbol{\theta}) = -2ig H_{\mathrm{XY}} + \frac{ig\chi(4-\mu)-1}{2n_0} \sum_j C_j(\boldsymbol{\theta})^2,
\label{eq:O_disrupt}
\end{equation}
where $C_j(\boldsymbol{\theta}) = \sum_k J_{jk}\cos(\theta_{jk})$, $n_0$ is the baseline polariton density, and $\chi = \partial n_0/\partial P $ is the basic coupling constant controlling the deviation. The first term of the perturbation operator $\mathcal{O}_{\mathrm{pert}}(\boldsymbol{\theta})$ originates from the polariton nonlinearity, which induces non-gradient oscillations that disrupt monotonic energy minimization. The second term is a four-spin interaction that introduces topological distortions to the energy landscape.

Physically, the coupling constant $\chi$ represents the number compressibility, quantifying the rate of change of the condensate density with respect to the applied non-resonant pump $P$. The magnitude of the perturbation operator is dictated by this parameter $\chi$. Below, we demonstrate that by treating the polariton density as a scaling parameter in an RG framework, $\chi$ can be driven to zero, rendering the perturbation term a irrelevant operator in the macroscopic limit.

\textbf{Renormalization-Group Analysis:--} Under a steady-state density rescaling transformation $n_j \to b n_j$ (with $b>1$), the local amplitude scales directly as $R_j \to \sqrt{b} R_j$. Because the rescaled amplitude must remain a valid solution of the steady-state equation, maintaining it via pump adjustment dictates that the compressibility transforms as $\chi \to b^{1-\mu/2}\chi$. The perturbation operator $\mathcal{O}_{\mathrm{pert}}$ remains bounded under this transformation, ensuring that the asymptotic fate of the perturbative correction $\chi \mathcal{O}_{\mathrm{pert}}$ is governed by the scaling of its coupling constant $\chi$ (see the Supplemental Material for the derivation). Equivalently, defining the continuous scale parameter $\ell=\ln b$ and the dimensionless ratio $\widetilde{\chi}(\ell) = \chi(e^\ell n_{\mathrm{ref}})/\chi(n_{\mathrm{ref}})$, we obtain the RG flow equation:
\begin{equation}
    \frac{d\widetilde{\chi}}{d\ell} = \beta(\mu) \,\widetilde{\chi},
\end{equation}
with the beta-function $\beta(\mu) \equiv d\ln\widetilde{\chi}/d\ell = 1-\mu/2$. Fig.~\ref{N4IntensityPhase}(b) illustrates the resulting RG flow of the compressibility with increasing density. For $\mu<2$, the coupling constant $\chi$ grows, whereas for $\mu>2$ the flow is driven toward the asymptotic scaling limit, $\chi=0$. At $\mu=2$, $\chi$ is marginal (density-independent). Thus, the scaling dimension $\beta$ of the compressibility changes sign at $\mu=2$, leading to the following far-reaching implications for the effective Hamiltonian:

(a) \textit{Compressible / Relevant regime ($\mu \le 2$):} For $\mu < 2$, the beta-function is positive ($\beta > 0$). Under macroscopic density scaling, the compressibility grows, making the perturbation operator $\mathcal{O}_{\mathrm{pert}}$ a \textit{relevant operators}. The effective dynamics becomes perturbed by these non-gradient disruptions, ensuring that the simulator fails to execute a pure gradient descent on the energy landscape. The conventional cubic-loss model ($\mu=2$) represents a marginal boundary where these disruptions persist at all scales, limiting optimization fidelity.

(b) \textit{Incompressible / Irrelevant regime ($\mu > 2$):} For higher-order nonlinear loss, the beta-function is negative ($\beta < 0$). Under density scaling, the flow is driven toward the attractive fixed point $\chi = 0$. Because its coupling constant vanishes asymptotically, the perturbation operator $\mathcal{O}_{\mathrm{pert}}$ becomes an \textit{irrelevant operator}. In this limit, both the four-body topological distortions and the non-gradient phase oscillations are ``integrated out" of the macroscopic physics. The normalized effective Hamiltonian recovers the pure, real-valued target system ($\mathcal{H}_{\mathrm{eff}} \to H_{\mathrm{XY}}$), giving the phase-only optimization.

(c) \textit{Coherent enhancement:} Stable amplitudes in the incompressible regime enable collective Josephson tunneling to bypass metastable traps and access deeper global minima. Unlike single-particle tunneling, this is a collective phenomenon inducing transitions between global phase configurations $|\boldsymbol{\theta}_A\rangle$ and $|\boldsymbol{\theta}_B\rangle$ representing nearby local minima. In this reduced two-state description, the real coupling $J^R_{jk}$ induces an effective off-diagonal matrix element:
\begin{equation}
 T_{AB}^{\mathrm{eff}} \propto \langle \boldsymbol{\theta}_A| \sum_{jk} J^R_{jk} (a_j^\dagger a_k + a_k^\dagger a_j ) |\boldsymbol{\theta}_B\rangle,
 \label{Tunneling}
\end{equation}
where $a_j^\dagger$ ($a_j$) are creation (annihilation) operators. Once polaritons transit toward a lower-energy basin through this term, the dissipative dynamics stabilize the new configuration. Thus, coherent coupling provides a quantum enhancement that drives convergence toward the true ground state.

In the following, we demonstrate that incompressibility enforces phase--amplitude decoupling, while coherent tunneling enhances optimization by escaping local minima, overcoming a fundamental limit of purely dissipative algorithms.

\begin{figure*}[t]
	\centering
	\includegraphics[width= 0.8\textwidth]{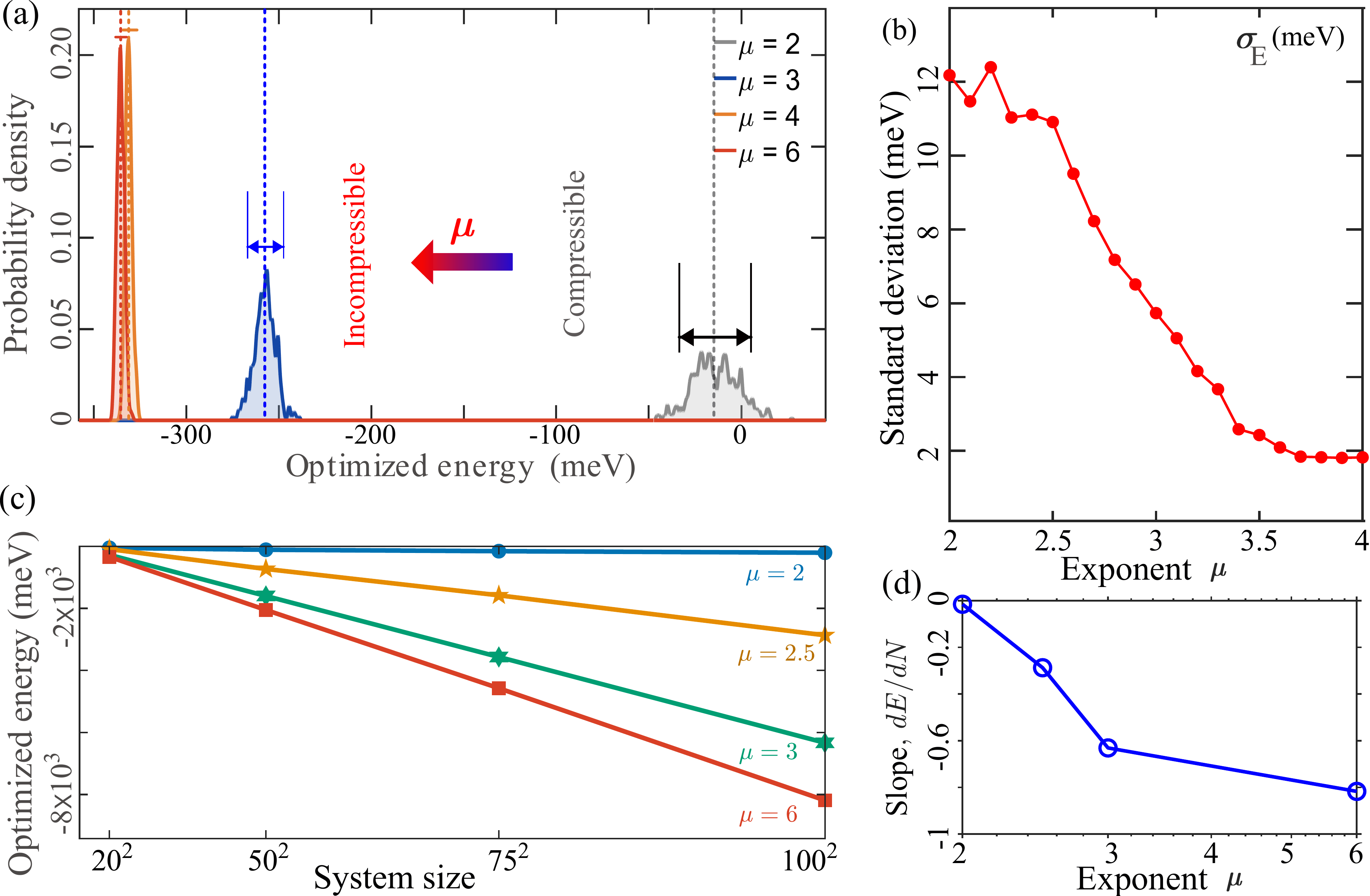}
    \caption{\textbf{The First Enhancement.} \textbf{(a)} Probability distributions of the optimized energies obtained from 500 independent simulations for an $N=20\times 20$ lattice at various exponents $\mu$. \textbf{(b)} The standard deviation of the optimized energy distribution as a function of $\mu$. It is reduced by a factor of six, demonstrating vastly improved stability. \textbf{(c)} Optimized energy as a function of system size $N$ for different values of $\mu$. The steeper scaling for larger $\mu$ demonstrates orders-of-magnitude improvements in the optimized energy for macroscopic systems. \textbf{(d)} The slope of the curves shown in (c).}
	\label{N4EnergyDistribution}
\end{figure*}

\begin{figure*}[t]
\centering
\includegraphics[width=1.6\columnwidth]{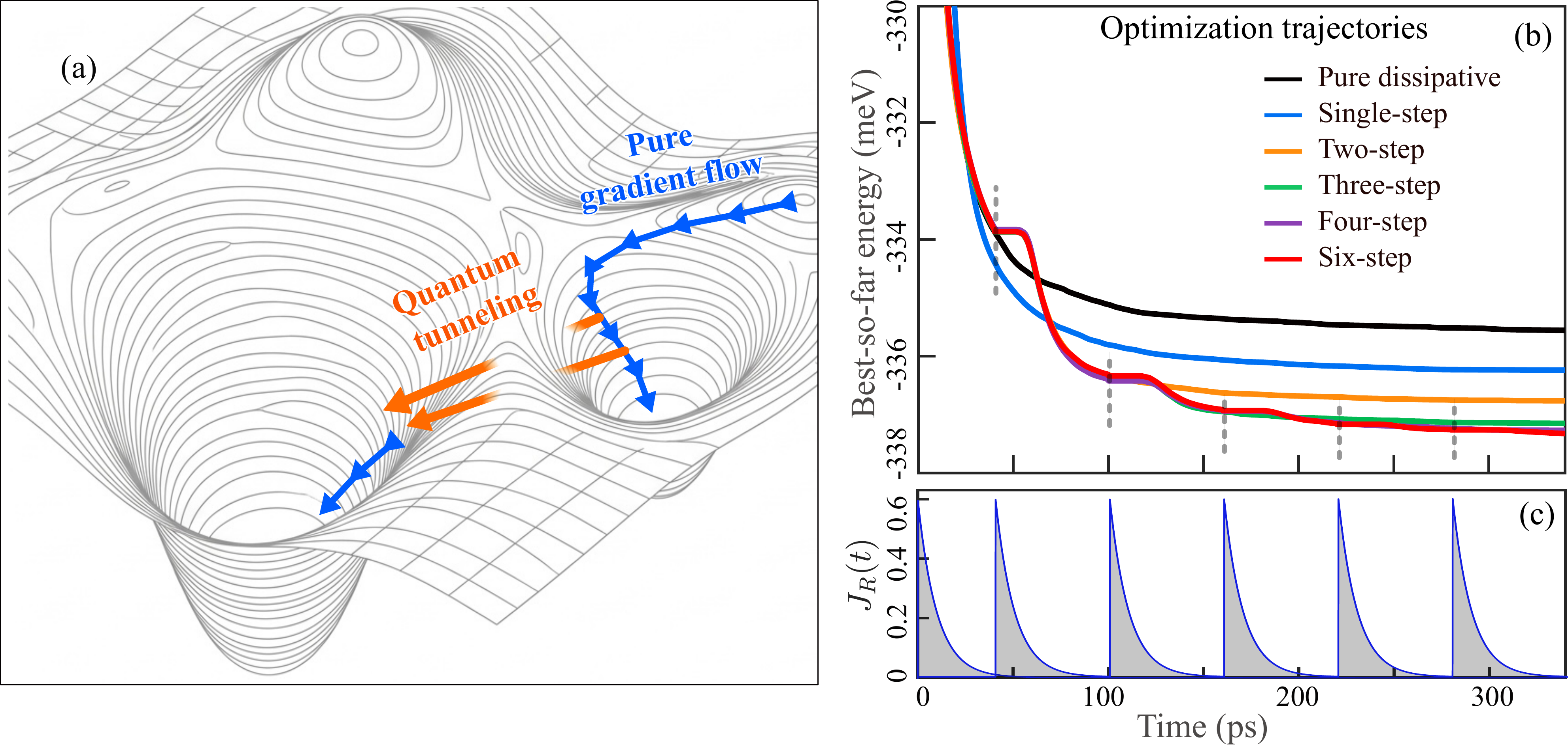}
\caption{\textbf{The Second Enhancement and a Multi-stage Protocol}. \textbf{(a)} Schematic representation of dissipative gradient descent into local minima and transverse exploration driven by coherent Josephson tunneling. \textbf{(b)} Optimization trajectories under pulsed coherent coupling in a large frustrated spin-glass system with size $N=400$. The best-so-far energy as a function of time for six protocols. The first enhancement (large-$\mu$ incompressible regime, $J^R_{jk}=0$) produces a substantial drop in the optimized energy. Each subsequent pulse of the coherent coupling further lowers the best-so-far energy, with the six-stage protocol achieving the lowest energy. \textbf{(c)} The corresponding time-dependent real coupling $J_R(t)$ for each protocol, shown on the same time axis for direct comparison. The pulsed protocols (two-, three-, four-, and six-stage) exhibit a periodic reset of $J_R$, whereas the always-on protocol decays monotonically and the pure dissipative protocol has $J_R=0$ throughout.}
\label{Quantumspeedup}
\end{figure*}

\textbf{First Enhancement:--} The transition from amplitude-fluctuation-dominated dynamics in the compressible regime to stabilized phase-only optimization in the large-$\mu$ regime is quantitatively summarized in Fig.~\ref{N4IntensityPhase}(c). For these simulations, we consider a two-dimensional square lattice with nearest-neighbor interactions, where the couplings are randomly distributed within $[-1, +1]$ meV. The resulting continuous spin-glass landscape provides a nonconvex benchmark; more generally, weighted XY minimization has the form of a unit-modulus quadratic program, which is NP-hard in general~\cite{8534389}. For each simulation, we record the minimum energy reached during the evolution. For $N=400$ (a $20 \times 20$ lattice), each marker represents the average of these minimum energies over 500 independent simulations with random initial conditions. Due to the steep computational scaling, each marker for $N>400$ corresponds to a single randomly initialized simulation. Across all investigated system sizes, the optimized energy drops sharply as $\mu$ increases, with the most pronounced reduction occurring during the transition toward the incompressible regime. For the largest simulated system size ($N=10,000$, a $100 \times 100$ lattice), we find a $40$-fold absolute energy reduction. For large $\mu$, the optimized energies gradually approach a size-dependent asymptotic minimum, demonstrating that suppressing phase--amplitude coupling fundamentally enhances the optimization dynamics.

For a closer look, Fig.~\ref{N4EnergyDistribution}(a) shows the probability distributions of the optimized energies obtained with random initial conditions for $N=400$. In the compressible regime ($\mu=2$), residual phase--amplitude coupling makes the effective phase dynamics strongly dependent on the local amplitude response. Consequently, simulations initialized from different random configurations converge to a wide range of metastable energies, producing a broad distribution of optimization outcomes. As $\mu$ increases, the distributions progressively shift toward lower energies and become substantially narrower, indicating simultaneous improvements in energy minimization and run-to-run reproducibility.

To quantify this narrowing of the statistical distribution, Fig.~\ref{N4EnergyDistribution}(b) shows the ensemble standard deviation of the optimized energy,
$ \sigma_E= \sqrt{ \langle E^2\rangle - \langle E\rangle^2 }$, where the averages are evaluated over 500 independent realizations. The large value of $\sigma_E\sim 12$ meV at $\mu=2$ reflects the strong sensitivity of the compressible dynamics to random initial conditions and the resulting convergence to widely distributed metastable basins. With increasing $\mu$, $\sigma_E$ decreases rapidly, showing that the optimized energies become increasingly reproducible. At $\mu=6$, the distribution is strongly concentrated within a sharply localized low-energy sector with $\sigma_E<2$ meV, representing a six-fold reduction. The combined downward shift of the energy distribution and suppression of $\sigma_E$ are consistent with the theoretical expectation that incompressibility reduces phase--amplitude cross-talk, stabilizes the effective phase-only dynamics, and improves the fidelity of the mapping to the target spin-glass Hamiltonian.

Fig.~\ref{N4EnergyDistribution}(c) further examines the scaling of the optimized energy with system size $N$ across different values of $\mu$. Because the true ground-state energy of a spin-glass model is an extensive quantity, a highly scalable solver should yield an average optimized energy that scales linearly with $N$. While this linear trend is generally observed, the solver's efficiency depends critically on the operational regime. The conventional compressible regime exhibits extremely poor scalability, capturing only a fraction of the extensive scaling with a shallow slope of $dE/dN \sim -0.015$ meV. In stark contrast, the incompressible regime successfully locates much deeper minima, demonstrating robust scaling with a steep slope of $dE/dN \sim -0.817$ meV. The dependence of this scaling slope on $\mu$ is summarized in Fig.~\ref{N4EnergyDistribution}(d). 
Consequently, for a macroscopic system of size $N=10^s$, the incompressible simulator achieves an absolute energy drop of $\sim 8 \times 10^{s-1}$ meV relative to the compressible baseline, where $s$ represents the logarithmic scale of the system. This diverging, orders-of-magnitude energy gap confirms that the optimization advantage induced by incompressibility persists at macroscopic scales.

\textbf{Second Enhancement:--} 
In polariton condensates, coherent Josephson coupling naturally arises due to the proximity effect via quantum tunneling. To incorporate this effect, we generalize Eq.~\eqref{coupled_dynamics} using the mean-field approximation of the Josephson term (Eq.~\eqref{Tunneling})~\cite{toebes2022dispersive,PhysRevLett.105.120403,kgz6-g8x3}. Consequently, an effective complex coupling emerges: $J_{jk} = J_{jk}^I + iJ_{jk}^R$, where $J_{jk}^I$ denotes the dissipative coupling responsible for loss minimization, and $J_{jk}^R$ denotes the Josephson coupling responsible for coherent exchange. With this complex coupling, the phase dynamics are described by:
\begin{eqnarray}
\hbar\dot{\theta}_j
&=&
-gR_j^2
-\sum_k\frac{R_k}{R_j}
\left[
J^I_{jk}\sin(\theta_{jk})
-
J^R_{jk}\cos(\theta_{jk})
\right],
\notag\\
\delta P_j
&\equiv&
\frac{2}{R_j}
\sum_k R_k
\left[
J^I_{jk}\cos(\theta_{jk})
+
J^R_{jk}\sin(\theta_{jk})
\right].
\label{phaseJR}
\end{eqnarray}
where the local pump correction $\delta P_j$ enters via Eq.~\eqref{coupled_dynamics}. Eq.~\eqref{phaseJR} explicitly distinguishes the roles of the two couplings. The dissipative component, $J^I_{jk}$, drives the system toward loss-landscape attractors, facilitating convergence to low-energy phase configurations. However, purely relaxation dynamics alone often become trapped in metastable local minima of the energy landscape. In contrast, the Josephson coupling, $J^R_{jk}$, introduces a $\cos(\theta_{kj})$ dependence that is mathematically orthogonal to the dissipative gradient $\sim \sin(\theta_{kj})$ [see Fig.~\ref{Quantumspeedup}(a)]. This contribution acts as a driving force, inducing an orthogonal exploration of the phase configuration.
Consequently, the dynamics are no longer purely gradient-like; nearby metastable configurations coherently mix, enabling effective inter-basin transfer.

Fig.~\ref{Quantumspeedup}(b) illustrates the effect of the coherent coupling using the same large-scale frustrated network considered above. The black line represents the best-so-far optimized energy obtained under purely dissipative coupling, while the colored curves represent the optimization trajectories enhanced by coherent tunneling effects. The purely dissipative dynamics initially lowers the energy rapidly but subsequently approaches a metastable plateau.

Fig.~\ref{Quantumspeedup}(c) shows the corresponding temporal modulation of the coherent coupling. In the purely dissipative protocol, $J_{jk}^{R}=0$ throughout the evolution.

\textbf{Multi-stage Protocol:--} To fully utilize coherent tunneling, we introduce a multi-stage protocol that applies the coherent coupling in successive pulses. In an $M$-stage protocol, the coherent term is reset at the onset of each stage $n$, evolving as:
\[
J_{jk}^{R}(t) = J_{jk}^{I}J_0e^{-(t-t_n)/\tau}, \qquad t\in[t_n,t_{n+1}).
\]
Each reset restores the coupling to its initial strength, reactivating transverse exploration just as the preceding dissipative relaxation begins to saturate. As shown in Fig.~\ref{Quantumspeedup}(b), these successive pulses yield distinct, stepwise reductions in the optimized energy, providing the system with repeated opportunities to escape metastable traps. 

We cap the protocol at six stages (an initial activation followed by five resets) as a practical compromise, balancing a linear increase in optimization time against diminishing energetic returns. These results demonstrate that repeatedly pulsing the coherent coupling systematically drives the system into deeper minima, strongly outperforming both purely dissipative relaxation and single-stage coherent evolution.

\begin{figure}[ht]
\centering
\includegraphics[width=1\columnwidth]{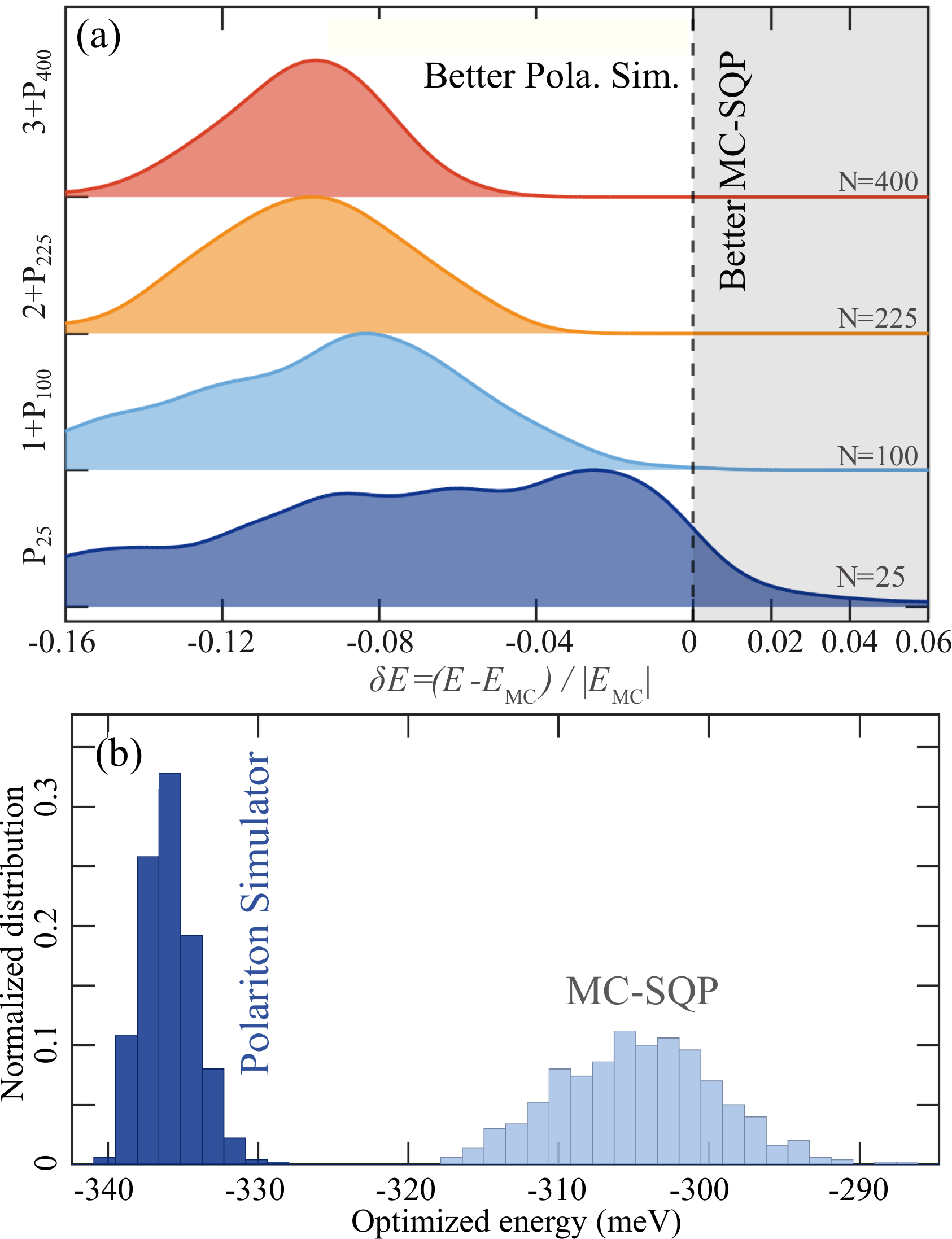}
\caption{\textbf{Scaling behavior of the polariton simulator in frustrated spin-glass systems.} \textbf{(a)} Distributions of the relative energy deviation, $\delta E = (E-E_{\mathrm{MC}})/|E_{\mathrm{MC}}|$, across various system sizes $N$, comparing the multi-stage polariton protocol with the classical MC--SQP benchmark. The progressive shift toward negative values demonstrates the simulator's growing systematic advantage in locating deeper energy minima as macroscopic complexity increases. \textbf{(b)} Direct comparison of the optimized energy distributions for $N=400$. The classical MC--SQP results exhibit a broad, higher-energy spread, whereas the polariton simulator yields a narrowed distribution at substantially lower energies. This contrast highlights the simulator's robust convergence capabilities in large, highly frustrated energy landscapes.}
\label{Scale}
\end{figure}

\textbf{Comparison with MC--SQP Optimization:--}
We finally benchmark the polariton simulator against a hybrid Monte Carlo--sequential quadratic programming (MC--SQP) approach~\cite{Boggs_Tolle_1995,doi:10.1137/S0036144504446096,fernández_solodov_2014,joshy2023sqp}. In each MC--SQP trial, a random phase configuration is generated to broadly sample different basins of the frustrated energy landscape, followed by strict local energy minimization via SQP. 

To quantify relative performance across different lattice sizes, we define the normalized energy deviation:
\begin{equation}
\delta E = \frac{E-E_{\mathrm{MC}}}{|E_{\mathrm{MC}}|},
\end{equation}
where $E$ and $E_{\mathrm{MC}}$ are the optimized energies obtained from the polariton simulator and the MC--SQP benchmark, respectively. Accordingly, $\delta E < 0$ indicates that the polariton simulator reaches a deeper energy minimum than the classical approach, whereas $\delta E > 0$ indicates the contrary.

Fig.~\ref{Scale}(a) shows the distributions of $\delta E$ for various lattice sizes. Even for the smallest systems considered, the distribution is heavily skewed toward $\delta E < 0$, indicating an immediate optimization advantage for the six-stage polariton simulator, though the classical solver occasionally finds a better minimum. As the system size and landscape complexity increase, the $\delta E$ distributions shift progressively toward more negative values, revealing a robust systematic advantage. For the largest network evaluated, the distribution lies entirely within the $\delta E < 0$ region, demonstrating that the incompressible polariton simulator outperforms the MC--SQP baseline in every sampled realization.

To explicitly illustrate this contrast, Fig.~\ref{Scale}(b) displays the raw optimized energy distributions for a representative system of size $N=400$. The MC--SQP results span a broad range of metastable energies, reflecting the inherent sensitivity of local minimizers to random initial conditions. Conversely, the polariton simulator yields a sharply concentrated distribution at significantly lower energies, with its ensemble average lying $30$ meV below the classical benchmark. This simultaneous downward shift and narrowed variance confirm that the multi-stage coherent protocol not only consistently unlocks deeper ground-state vicinities, but also guarantees reproducible optimization outcomes.

\section*{Conclusion}
In this work, we identified and resolved major challenges to scaling polariton-based spin simulators. We showed that conventional simulators are limited by phase--amplitude coupling, which degrades their fidelity on macroscopic networks. Accordingly, by driving the condensates toward an incompressible RG fixed point via higher-order nonlinear loss, these disruptive amplitude fluctuations are strongly suppressed. This stabilization is essential for the programmability of the polariton simulator to faithfully reproduce the target spin dynamics at macroscopic scales.

The practical impact of this stabilization becomes striking when evaluating highly frustrated spin-glass models. We found that conventional solvers fail to capture the true scaling of the ground state, meaning their optimization efficiency plummets as the network grows. In contrast, the incompressible regime successfully recovers this fundamental linear scaling. As a result, the performance gap between the two approaches widens dramatically with system size. 

To overcome the metastable trapping characteristic of such NP-hard energy landscapes, we introduced a weak Hermitian pulsed coupling operating alongside the dissipative evolution. This coherent coupling induces a transverse exploration of the phase space---orthogonal to the dissipative gradient---driving collective coherent tunneling and inter-basin mixing. By integrating this mechanism into a multi-stage protocol of successive decaying pulses, the system repeatedly escapes local minima to locate vastly deeper energy sectors. Extensive benchmark comparisons against a classical MC--SQP baseline confirm a robust, systematically widening scalability advantage: for the largest evaluated networks, the incompressible polariton simulator outperformed the classical local minimizer in every sampled realization.

These results provide a unified optimization architecture wherein nonlinear loss guarantees macroscopic phase stability and coherent Josephson dynamics unlock powerful exploratory tunneling. By successfully navigating the continuous, uncountably infinite search space of randomly coupled XY spin glasses, our work establishes driven--dissipative condensate networks as robust, highly scalable analog platforms for complex  energy minimization.

While preparing this manuscript, we became aware of related work~\cite{privatecomm} that independently identifies the limitations of conventional XY simulators for phase encoding under standard nonlinearity.

\section*{Acknowledgments} 
S.G. acknowledges funding supports from the National Natural Science Foundation of China (Grant No. 12274034), the Guangdong Province Foreign Experts Project (Flexible Talent Introduction) Fund (No. 2025TB0050), Shenzhen Specially Appointed Positions (No. 2026TC0047), and University Development Fund (No. UDF01003913).

\section{Supplementary Information}

Conventional models of nonlinear loss in exciton-polariton condensates typically predict a linear growth of the condensate population with respect to the incoherent pump power. However, fundamental physical constraints and recent experimental observations reveal sublinear growth dynamics, pointing to the presence of higher-order nonlinear loss mechanisms. Below, we provide a detailed physical analysis of these factors. Furthermore, we extract the effective exponent $\mu$ from experimentally reported power-dependence curves, finding that it can span a broad range from approximately $1$ to $20$. This confirms that the incompressible regime ($\mu > 2$) underlying our proposed optimization mechanism is readily accessible in current experimental platforms.

\subsection{Conventional Linear Growth Models}
We begin by reviewing the standard open-dissipative Gross--Pitaevskii model for a nonresonantly pumped polariton condensate coupled to an incoherent excitonic reservoir \cite{wouters2007excitations,RevModPhys.85.299}. For each condensate site, the local dynamics are governed by:
\begin{align}
i\hbar\dot{\psi}_{j}
&=
\hat{H}_{j}^{(0)}\psi_{j}
+g_{c}|\psi_{j}|^{2}\psi_{j} +\frac{i\hbar}{2}
\left(Rn_{R,j}-\gamma_{c}\right)\psi_{j},
\label{eq:reservoir_gpe}
\\
\dot{n}_{R,j}
&=
P_{j}-\gamma_{R}n_{R,j}
-Rn_{R,j}|\psi_{j}|^{2},
\label{eq:reservoir_rate}
\end{align}
where \(n_{R,j}\) is the reservoir density, \(P_j\) is the nonresonant pump rate, \(R\) is the stimulated scattering coefficient, and \(\gamma_c\) and \(\gamma_R\) are the condensate and reservoir decay rates, respectively. The operator \(\hat{H}_{j}^{(0)}\) encapsulates the conservative single-site contributions.

Because the excitonic reservoir typically relaxes on a much faster timescale than the polariton condensate, this separation of timescales allows for the adiabatic elimination of the reservoir dynamics. Setting \(\dot n_{R,j}\simeq 0\) yields:
\begin{equation}
n_{R,j}
\simeq
\frac{P_j}{\gamma_R+R|\psi_j|^2}.
\label{eq:adiabatic_reservoir}
\end{equation}
Introducing the local condensate density \(n_j=|\psi_j|^2\), the resulting density-dependent effective net gain is:
\begin{equation}
G_j(n_j)
=
\frac{RP_j}{\gamma_R+Rn_j}-\gamma_c.
\label{eq:reservoir_gain}
\end{equation}
The steady-state condition, \(G_j(n_j)=0\), trivially reduces to:
\begin{equation}
n_j
=
\frac{P_j-P_{\mathrm{th}}}{\gamma_c},
\qquad
P_{\mathrm{th}}
=
\frac{\gamma_c\gamma_R}{R}.
\label{eq:standard_density}
\end{equation}
As expressed in Eq.~\eqref{eq:standard_density}, this conventional model predicts a linear, unbounded growth of the condensate density as a function of the incoherent pump $P_j$ above the condensation threshold $P_{\mathrm{th}}$.

From an effective theory perspective, the same effect can be viewed as the following effective nonlinear loss:
\begin{equation}
i\hbar\dot{\psi}_{j}
= \hat{H}_{j}^{(0)}\psi_{j}
+g_{c}|\psi_{j}|^{2}\psi_{j} +\frac{i\hbar}{2} [P_j-2\alpha|\psi_j|^\mu-\gamma]\psi_{j}  ,
\label{eq:power_loss}
\end{equation}
the conventional reservoir model corresponds to \(\mu=2\), leading to the steady state density:
\begin{equation}
n_j = \frac{P_j-\gamma}{2\alpha}
\label{eq:standard_density2}
\end{equation}
Therefore, both the conventional effective nonlinear loss term and the two-component rate equations leads to the same effective linear growth (Eq.~\eqref{eq:standard_density} or \eqref{eq:standard_density2}) of the condensate density with external incoherent pump $P_j$.

\subsection{Sublinear Growth and Effective Higher-Order Nonlinear Loss}
At elevated excitation densities, the conventional reservoir description becomes insufficient because additional density-dependent processes fundamentally alter the balance between gain and loss. For instance, phase-space filling reduces the excitonic oscillator strength as available electron and hole states are progressively occupied \cite{PhysRevB.49.16565}. Accordingly, intensity-dependent microcavity measurements have revealed the bleaching of the strong-coupling response and a pronounced reduction in oscillator strength at high excitation densities \cite{PhysRevB.52.7810}. Related nonlinearities have also been observed through exciton blockade \cite{gu2021enhanced}, oscillator-strength quenching \cite{r8m8-hckx}, and density-dependent reductions of the Rabi splitting in strongly driven microcavities \cite{dstp-cbcf}.

Collectively, these microscopic effects dictate that the coherent polariton population cannot continue to scale linearly with pump power. Indeed, sublinear population growth has been widely reported across several polariton platforms, including confined ZnO condensates \cite{PhysRevB.91.165423}, room-temperature perovskite microcavities \cite{10.1021/acsphotonics.6c00428}, and in measurements separating coherent and thermal polariton populations \cite{PRXQuantum.2.030320}. Although the specific experimental observables and microscopic details may vary among these platforms, they universally manifest as a diminishing growth rate of the condensate density as a function of the applied pump power.
 
In general, the condensate growth rate can be quantified by the logarithmic population response:
\begin{equation}
\beta
\equiv
\frac{d\ln n_j}{d\ln\Delta P_j}
=
\frac{2}{\mu},
\label{eq:growth_exponent}
\end{equation}
where $\Delta P_j = P_j - P_{\mathrm{th}}$ is the net pump rate above threshold. For the conventional cubic-loss case (\(\mu=2\)), one recovers \(\beta=1\), corresponding to linear population growth. For any finite \(\mu>2\), the response becomes sublinear (\(0<\beta<1\)).

Consequently, for any general experimental population curve, a local effective nonlinear-loss exponent can be directly extracted from its logarithmic slope:
\begin{equation}
\mu = 2 \frac{d\ln(P-P_{\mathrm{th}})}{d\ln n}.
\label{eq:effective_mu}
\end{equation}
This formulation elegantly characterizes sublinear growth without necessitating a specific microscopic mechanism. Within our framework, \(\mu\) serves as a compact, effective parameter that encodes the cumulative impact of the reduced response at elevated pumping.

Applying Eq.~\eqref{eq:effective_mu} to the experimental data reported in Refs.~\cite{PhysRevB.91.165423, 10.1021/acsphotonics.6c00428, PRXQuantum.2.030320}, we find that the effective value of $\mu$ consistently exceeds $2$ in the high-power regime. Depending on the specific material platform and excitation intensity, our analysis reveals that $\mu$ spans a broad range, from approximately $1$ to $20$. This confirms that the large-$\mu$ incompressible regime leveraged by our optimization protocol is fundamentally justified and readily accessible in current experimental setups.

\subsection{Phase-Only Effective Theory}
\label{sec:phase_only_effective_functional}

To formally quantify how amplitude fluctuations corrupt the intended phase dynamics, we derive an effective phase-only Hamiltonian for the polariton network. We start from the driven-dissipative Gross--Pitaevskii equation (dGPE):
\begin{equation}
\hbar\dot{\psi}_j
=
\sum_k J_{jk}\psi_k
+\left(\frac{P-\gamma}{2}-\alpha|\psi_j|^\mu\right)\psi_j
-i g|\psi_j|^2\psi_j .
\label{eq:dgpe_start}
\end{equation}

\subsubsection{Formal Imaginary-Time Generator}
Because analog simulators seek to minimize a cost function through relaxation, it is highly instructive to cast the dGPE into a formal imaginary-time evolution form:
\begin{equation}
\hbar\dot{\psi}_j
=
-\frac{\partial H_{\mathrm{form}}}{\partial\psi_j^*},
\label{eq:imaginary_time_form}
\end{equation}
where $\psi_j$ and $\psi_j^*$ are treated as independent variables when taking the derivative. By integrating the right-hand side of Eq.~\eqref{eq:dgpe_start}, we construct the corresponding formal generator:
\begin{align}
H_{\mathrm{form}}(\psi,\psi^*)
={}&
-\sum_{j,k}J_{jk}\psi_j^*\psi_k
-\frac{P-\gamma}{2}\sum_j|\psi_j|^2
\nonumber\\
&+\frac{2\alpha}{\mu+2}\sum_j|\psi_j|^{\mu+2}
+\frac{ig}{2}\sum_j|\psi_j|^4 .
\label{eq:H_formal_complex}
\end{align}
Applying the derivative $-\partial/\partial\psi_j^*$ exactly recovers Eq.~\eqref{eq:dgpe_start}. To separate the phase and amplitude degrees of freedom, we apply the Madelung transformation, $\psi_j=R_je^{i\theta_j}$. Defining the relative phase difference as $\Delta_{kj}=\theta_k-\theta_j$ and assuming real, symmetric couplings ($J_{jk}=J_{kj}$), the formal generator becomes:
\begin{align}
H_{\mathrm{form}}(R,\theta)
={}&
-\sum_{j,k}J_{jk}R_jR_k\cos(\Delta_{kj})
-\frac{P-\gamma}{2}\sum_jR_j^2
\nonumber\\
&+\frac{2\alpha}{\mu+2}\sum_jR_j^{\mu+2}
+\frac{ig}{2}\sum_jR_j^4 .
\label{eq:H_formal_Rtheta}
\end{align}

\subsubsection{Steady-State Amplitude}
The amplitude dynamics dictated by Eq.~\eqref{eq:dgpe_start} are given by:
\begin{equation}
\hbar\dot R_j
=
\frac{P-\gamma}{2}R_j
-\alpha R_j^{\mu+1}
+\sum_kJ_{jk}R_k\cos(\theta_{kj}).
\label{eq:amplitude_equation}
\end{equation}
In the absence of network couplings ($J_{jk}=0$), the system relaxes to a uniform steady-state amplitude $R_0$ and baseline density $n_0$:
\begin{equation}
R_0=\left(\frac{P-\gamma}{2\alpha}\right)^{1/\mu},
\qquad
n_0=R_0^2
=\left(\frac{P-\gamma}{2\alpha}\right)^{2/\mu}.
\label{eq:R0_n0}
\end{equation}
When the network couplings are introduced, the amplitudes become spatially nonuniform. To capture this, we define the local phase-induced force $C_j(\boldsymbol{\theta})$:
\begin{equation}
C_j(\boldsymbol{\theta})
=\sum_kJ_{jk}\cos(\theta_k-\theta_j),
\qquad
\chi \simeq \frac{\partial n_j}{\partial P}
\label{eq:C_chi}
\end{equation}
Assuming the coupling-induced perturbations are weak relative to the uniform gain ($|C_j|\ll \mu(P-\gamma)/2$), we can perturbatively solve the steady-state amplitude equation ($\hbar\dot R_j \approx 0$). This yields the steady-state density:
\begin{equation}
n_j^{\mathrm{ss}}
\simeq
n_0+2\chi C_j
+\frac{\chi^2}{n_0}
\left[
2\sum_kJ_{jk}\cos(\theta_k-\theta_j)C_k
-\mu C_j^2
\right]
\label{eq:n_ss_second_order}
\end{equation}
where the parameter $\chi$ appears as the effective parameter that quantifies the amplitude (density) fluctuations due to its coupling with the phase. We identify that 
\begin{equation}
\chi = \frac{\partial n_0^\mathrm{ss} }{\partial P} \equiv \lim_{\gamma t \to \infty} 2 R_0 \frac{\partial R_0 }{\partial P} 
\end{equation}

\subsubsection{The Effective Phase Hamiltonian}
At long times, the driven-dissipative system approaches a near-steady-state condition where gain and loss are balanced. Because the amplitudes relax much faster than the global phase configuration, we can construct an effective phase-only Hamiltonian by substituting the amplitudes directly into the formal generator:
\begin{equation}
\bar n_j(\boldsymbol{\theta})
\equiv n_j^{\mathrm{ss}}(\boldsymbol{\theta}),
\qquad
\bar R_j(\boldsymbol{\theta})
=\sqrt{\bar n_j(\boldsymbol{\theta})}.
\label{eq:bar_n_R}
\end{equation}
Substituting Eq.~\eqref{eq:bar_n_R} into Eq.~\eqref{eq:H_formal_Rtheta} restricts the dynamics to a phase-only manifold, yielding the effective functional:
\begin{align}
H_{\mathrm{eff}}(\boldsymbol{\theta})
={}&
-\sum_{j,k}J_{jk}
\sqrt{\bar n_j(\boldsymbol{\theta})\bar n_k(\boldsymbol{\theta})}
\cos(\theta_k-\theta_j)
\nonumber\\
&-\frac{P-\gamma}{2}\sum_j\bar n_j(\boldsymbol{\theta})
\nonumber\\
&+\frac{2\alpha}{\mu+2}
\sum_j\bar n_j(\boldsymbol{\theta})^{1+\mu/2}
+\frac{ig}{2}\sum_j\bar n_j(\boldsymbol{\theta})^2 .
\label{eq:H_eff_exact_substitution}
\end{align}

To explicitly reveal the impact of the amplitude fluctuations on the optimization landscape, we define the pure target XY Hamiltonian:
\begin{equation}
H_{\mathrm{XY}}
=-\sum_{j<k}J_{jk}\cos(\theta_k-\theta_j).
\label{eq:H_XY_def_final}
\end{equation}
By expanding $H_{\mathrm{eff}}(\boldsymbol{\theta})$ using Eq.~\eqref{eq:n_ss_second_order} and retaining terms up to $O(J^2)$, we obtain:
\begin{align}
H_{\mathrm{eff}}(\boldsymbol{\theta})
={}&h_0+2n_0H_{\mathrm{XY}}
-4ig\chi n_0H_{\mathrm{XY}}
\nonumber\\
&+\left[-\chi+i(4-\mu)g\chi^2\right]
\sum_j C_j(\boldsymbol{\theta})^2
+O(J^3),
\label{eq:H_eff_expanded}
\end{align}
where $h_0$ is an inconsequential phase-independent constant:
\begin{equation}
h_0
=
N\left[
-\frac{\alpha\mu}{\mu+2}n_0^{1+\mu/2}
+\frac{ig}{2}n_0^2
\right].
\label{eq:h0}
\end{equation}

Equation~\eqref{eq:H_eff_expanded} elegantly exposes the physical failure mode of conventional polariton simulators. While the term proportional to $H_{\mathrm{XY}}$ represents the intended target dynamics, the term proportional to $\sum_j C_j(\boldsymbol{\theta})^2$ represents a amplitude-feedback correction. Because each $C_j$ is a sum of weighted cosines, this correction injects complex four-body spin interactions into the energy landscape. 

The magnitude of this disruptive cross-talk is dictated by the compressibility $\chi$. In the incompressible scaling limit ($\chi \to 0$), all amplitude-dependent corrections vanish:
\begin{equation}
H_{\mathrm{eff}}
\longrightarrow h_0+2n_0H_{\mathrm{XY}}.
\end{equation}
After dropping the phase-independent constant $h_0$ and dividing by the positive scaling factor $2n_0$, the system perfectly recovers the pure target XY Hamiltonian. Thus, driving the system into the large-$\mu$ incompressible regime is a strict mathematical requirement for high-fidelity XY optimization.

\subsection{Renormalization-Group Flow and Irrelevant Operators}
\label{sec:rg_flow}

Equation~\eqref{eq:H_eff_expanded} reveals that the physical polariton network does not minimize the pure, real-valued target Hamiltonian. Instead, it evolves on an effective, non-Hermitian landscape corrupted by higher-order phase interactions and imaginary oscillatory forces. To establish the macroscopic scalability of the simulator, we cast this effective Hamiltonian into a renormalization-group (RG) framework. 

Ignoring the phase-independent constant $h_0$ and factoring out the dominant $2n_0$ scaling of the pure target dynamics, the normalized effective phase Hamiltonian takes the form:
\begin{equation}
{\mathcal{H}}_{\mathrm{eff}}(\boldsymbol{\theta}) = \frac{H_{\mathrm{eff}} - h_0}{2n_0} = H_{\mathrm{XY}} + \chi \mathcal{O}_{\mathrm{disrupt}},
\label{eq:H_tilde}
\end{equation}
where we identify $\chi$ as the fundamental coupling constant controlling all disruptive dynamics. The corresponding disruption operator $\mathcal{O}_{\mathrm{disrupt}}$ is given by:
\begin{equation}
\mathcal{O}_{\mathrm{disrupt}} = -2ig H_{\mathrm{XY}} + \left[ -\frac{1}{2n_0} + i\frac{(4-\mu)g\chi}{2n_0} \right] \sum_j C_j(\boldsymbol{\theta})^2.
\label{eq:O_disrupt}
\end{equation}

The operator $\mathcal{O}_{\mathrm{disrupt}}$ contains two distinct failure modes for analog optimization. The second term ($\propto C_j^2$) introduces complex four-body spin interactions that topologically distort the energy landscape, creating false minima. The first term ($-2ig H_{\mathrm{XY}}$) introduces an imaginary, non-Hermitian contribution to the target Hamiltonian. In the time-evolution dynamics ($\hbar\dot{\theta}_j = -\partial {\mathcal{H}}_{\mathrm{eff}} / \partial \theta_j$), this imaginary component generates phase oscillations, entirely derailing the monotonic gradient descent required to reliably locate the ground state.

\subsubsection{Density Renormalization-Group Flow}

To analyze the macroscopic behavior of the effective phase theory, we consider a density rescaling transformation ${n}_0 \to b {n}_0$ (with $b > 1$). Because $H_{\mathrm{XY}}$ and $  C_j(\boldsymbol{\theta})$ are density independent, they remain invariant while the amplitudes rescale directly:
\begin{align}
  H_{\mathrm{XY}} &\to H_{\mathrm{XY}}, \notag \\
  C_j(\boldsymbol{\theta}) &\to C_j(\boldsymbol{\theta}), \notag \\
  R_j &\to \sqrt{b} R_j.
\end{align}

Determining the scaling transformation of the compressibility $\chi$ requires careful consideration. For the rescaled baseline density $b n_0$ to remain a valid physical state, it must continue to satisfy the steady-state condition dictated by Eq.~\eqref{eq:amplitude_equation}. Physically, this is achieved by adjusting the applied pump from $P$ to a rescaled value $P_b$ such that the new steady-state density is $b n_0$. Applying the unperturbed steady-state relation to both scales and subtracting yields:
\begin{align}
\frac{P_b-P}{2}
&=\alpha\left[(bn_0)^{\mu/2}-n_0^{\mu/2}\right], \notag \\
P_b-\gamma
&=P-\gamma+2\alpha n_0^{\mu/2}
\left(b^{\mu/2}-1\right).
\label{eq:pump_rescaling}
\end{align}
In the weak-coupling regime ($J_{jk}$ small), the baseline steady-state relation is $P-\gamma=2\alpha n_0^{\mu/2}$, which simplifies Eq.~\eqref{eq:pump_rescaling} to:
\begin{equation}
P_b-\gamma = b^{\mu/2}(P-\gamma).
\end{equation}
Evaluating the compressibility along this steady-state manifold yields the exact scaling transformation:
\begin{equation}
\chi \to b^{\,1-\mu/2}\chi.
\label{eq:chi_scaling}
\end{equation}
Thus, for a given network, scaling the steady-state density by a factor $b$ via pump adjustment rescales the macroscopic compressibility by $b^{1-\mu/2}$.

Next, we examine the internal scaling of the perturbation operator $\mathcal{O}_{\mathrm{pert}}$ itself. Substituting $n_0 \to b n_0$ and Eq.~\eqref{eq:chi_scaling} into Eq.~\eqref{eq:O_disrupt} gives:
\begin{align}
\mathcal{O}_{\mathrm{pert}} \to{}& -2igH_{\mathrm{XY}} -\frac{b^{-1}}{2n_0} \sum_j C_j(\boldsymbol{\theta})^2 \notag \\
&+ i\frac{b^{-\mu/2}}{2n_0}g\chi (4-\mu)\sum_j C_j(\boldsymbol{\theta})^2.
\end{align}
Crucially, for any physical exponent $\mu > 0$, none of the internal terms within $\mathcal{O}_{\mathrm{pert}}$ diverge under density rescaling ($b > 1$): the first term is scale-invariant ($\sim b^0$), the second decays as $\sim b^{-1}$, and the third decays as $\sim b^{-\mu/2}$. Because the perturbation operator is bounded and does not grow intrinsically, the asymptotic fate of the full perturbation $\chi \mathcal{O}_{\mathrm{pert}}$ is governed entirely by the RG flow of its prefactor $\chi$.

Defining the continuous logarithmic scale parameter $\ell = \ln(n_0 / n_{\mathrm{ref}}) = \ln b$, the density rescaling transformation yields the exact RG flow equation for the running coupling constant $\chi(\ell)$:
\begin{equation}
\frac{d\chi}{d\ell} = \beta(\chi) = \left(1 - \frac{\mu}{2}\right) \chi.
\label{eq:beta_function}
\end{equation}

This flow equation allows us to formally classify the perturbation operator $\mathcal{O}_{\mathrm{pert}}$ according to standard RG terminology:

\begin{itemize}
    \item \textbf{Compressible / Relevant Regime ($\mu < 2$):} The beta-function is positive ($\beta > 0$). Under macroscopic density scaling, the coupling constant $\chi$ grows, rendering the non-Hermitian oscillations and four-body distortions \textit{relevant perturbations}. The effective Hamiltonian becomes heavily dominated by these terms, ensuring that the simulator fails to execute a pure gradient descent on the XY landscape. The conventional cubic-loss model ($\mu=2$) represents a marginal boundary ($\beta = 0$) where the leading-order oscillatory disruption persists at all scales.
    
    \item \textbf{Incompressible / Irrelevant Regime ($\mu > 2$):} The beta-function is negative ($\beta < 0$). Under density scaling ($b \to \infty$), the flow is driven deterministically toward the attractive fixed point $\chi^* = 0$. Because its coupling constant vanishes asymptotically, the entire perturbation operator $\mathcal{O}_{\mathrm{pert}}$ becomes a \textit{irrelevant operator}.
\end{itemize}

At this incompressible asymptote ($\chi \to 0$), both the four-body topological distortions and the non-Hermitian phase oscillations are rigorously suppressed. The normalized effective Hamiltonian perfectly recovers the pure, real-valued target system:
\begin{equation}
\widetilde{\mathcal{H}}_{\mathrm{eff}}(\boldsymbol{\theta}) \xrightarrow[\mathrm{RG\ flow}]{\mu > 2} H_{\mathrm{XY}}.
\end{equation}
Consequently, driving the system into the large-$\mu$ incompressible scaling limit is a strict mathematical requirement for the polariton network to execute stable, high-fidelity gradient descent on the macroscopic XY landscape.

\subsection{Robustness Across Coupling Topologies}

The calculations presented in the main text focus on a two-dimensional square lattice with nearest-neighbor coupling, for which each bulk site has a coordination number \(z=4\). To determine whether the optimization enhancement induced by higher-order nonlinear loss depends on this particular geometry, we consider two additional coupling topologies: a one-dimensional nearest-neighbor ring with \(z=2\) and an all-to-all-coupled network with \(z=N-1\). These two cases probe, respectively, a sparse low-connectivity limit and a dense connectivity limit.

For both topologies, the dissipative coupling coefficients \(J_{jk}^{I}\) are sampled from a uniform distribution over \([-1,1]\), symmetrized, and assigned zero diagonal elements. The coherent coupling is set to \(J_{jk}^{R}=0\) in order to isolate the enhancement arising from higher-order nonlinear loss. For each system size, the same coupling realization and initial condition are retained when varying \(\mu\), so that the resulting energy variation primarily reflects the change in condensate compressibility. The system is evolved to \(t_{\mathrm{f}}=100~\mathrm{ps}\), and the final XY energy is recorded as the optimized energy \(E_{\min}\).

\begin{figure}[t]
\centering
\includegraphics[width=1\columnwidth]{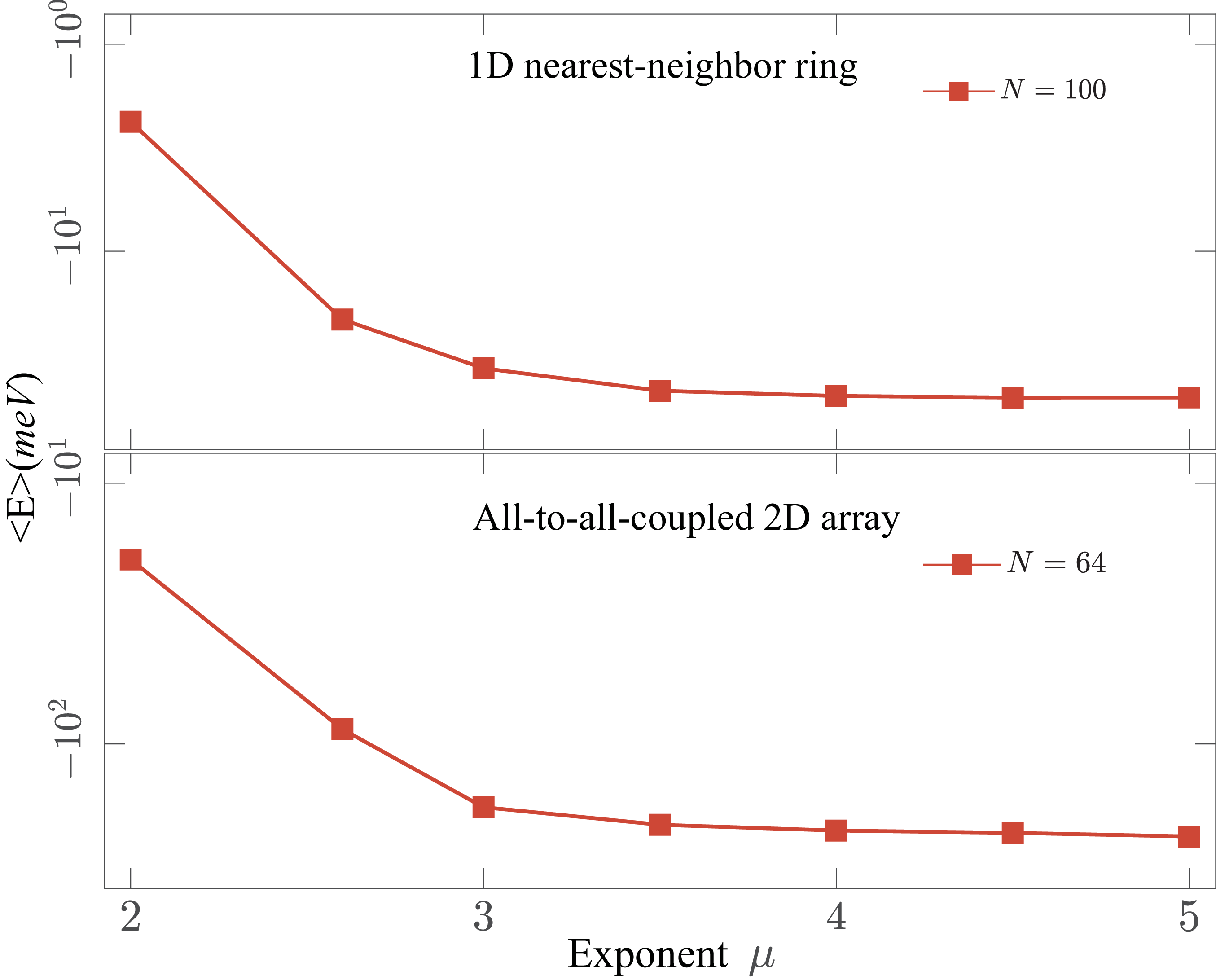}
\caption{
Optimized XY energy \(E_{\min}\) as a function of the nonlinear-loss exponent \(\mu\) for coupling topologies beyond the two-dimensional nearest-neighbor lattice considered in the main text. The upper panel shows one-dimensional nearest-neighbor rings with \(N=100\) and \(N=400\), while the lower panel shows all-to-all-coupled networks with the same system sizes. The energy magnitude is displayed on a logarithmic scale, with the corresponding negative-energy values retained as the axis labels.
}
\label{fig:supp_topology}
\end{figure}

As shown in the upper panel of Fig.~\ref{fig:supp_topology}, increasing \(\mu\) produces a pronounced reduction in the optimized energy of the one-dimensional rings for both system sizes. The energy decreases most rapidly between the conventional compressible regime at \(\mu=2\) and the crossover toward the incompressible regime, before approaching a low-energy plateau around \(\mu\simeq4\)--\(5\). At \(\mu=5\), the corresponding energies per site are approximately \(E_{\min}/N=-0.509\) and \(-0.505\) for \(N=100\) and \(N=400\), respectively. Their close agreement indicates that the improvement remains effective as the ring size increases and is not a finite-size feature of the smaller system.

The all-to-all-coupled networks exhibit the same overall dependence on \(\mu\), as shown in the lower panel of Fig.~\ref{fig:supp_topology}. Despite the substantially denser coupling structure, increasing the nonlinear-loss exponent drives the system toward markedly lower optimized energies, with the strongest improvement again occurring upon leaving the \(\mu=2\) regime. The comparatively small variations among the largest-\(\mu\) points reflect the stochastic character of individual optimization trajectories after the dynamics has entered the low-energy sector.

The absolute energies in the two panels should not be compared directly because the number of active bonds scales as \(O(N)\) for the nearest-neighbor ring but as \(O(N^{2})\) for the all-to-all-coupled network. The relevant observation is instead their consistent dependence on \(\mu\). Together with the two-dimensional nearest-neighbor results in the main text, these calculations demonstrate that the improvement produced by higher-order nonlinear loss persists across networks with widely different dimensionalities and connectivities. The incompressibility-induced enhancement is therefore not tied to a specific lattice geometry, but provides a robust mechanism for improving the mapping of driven--dissipative polariton networks onto frustrated XY Hamiltonians.

\bibliographystyle{unsrt}
\bibliography{ref}
\end{document}